\documentclass[aps,prl,preprint,showpacs,groupedaddress]{revtex4}
\usepackage {amssymb}
\usepackage {amsmath}
\usepackage {graphicx}
\usepackage {longtable}

\begin{document}

\title{Stabilization of nonlinear plasma waves by radiation}

\author{Fedor V.Prigara}
\affiliation{Institute of Microelectronics and Informatics,
Russian Academy of Sciences,\\ 21 Universitetskaya, Yaroslavl
150007, Russia}
\email{fprigara@imras.yar.ru}

\date{\today}

\begin{abstract}
It is shown that the interaction of a plasma with thermal
radiation leads to the stabilization of both periodic and solitary
nonlinear plasma waves. The stabilized periodic nonlinear plasma
waves were indeed observed in the experiment carried out by Looney
and Brown. There are many branches of radiation-induced solitary
waves in plasmas, corresponding in particular to the density waves
in hot plasmas of accretion disks.
\end{abstract}


\maketitle

It is known that longitudinal plasma waves are subject to the linear and
nonlinear Landau damping [1,2]. On the other hand, electron plasma waves are
unstable with respect to the modulation instability and formation of
envelope solitons. However, in a plasma without a magnetic field, plane
solitons should be also unstable due to the transverse modulation, and
therefore they cannot exist, according to the theoretical predictions [3].
The theories of Langmuir wave collapse and Landau damping do not take into
account the interaction of a plasma with thermal radiation. Here we show
that the interaction of a plasma with radiation leads to the stabilization
of both periodic and solitary electron plasma waves. Note that there is no
direct experimental evidence of the existing of Langmuir wave collapse [3].

In the experiment carried out by Looney and Brown [4], a beam of several
hundred \textit{eV} electrons, injected into the plasma of a dc discharge
from an auxiliary electron gun, excited oscillations in the plasma at the
plasma electron oscillation frequency. However, Looney and Brown have
observed that the frequency of oscillation varies in a discontinuous manner
with the electron density $n_{e} $ which is proportional to the gun
discharge current. The frequency remained nearly constant over small
variations of the electron density and then jumped abruptly to a new
frequency. Along one of these frequency plateaus, the intensity of the
oscillation rose to a maximum and then fell to a much lower value before a
jump to a new frequency occured. The frequency corresponding to the
strongest signal for each frequency range was consistent with the Tonks -
Langmuir relation

\begin{equation}
\label{eq1}
\omega _{pe} = \left( {4\pi n_{e} e^{2}/m_{e}}  \right)^{1/2},
\end{equation}

\noindent
where $m_{e} $ is the mass of an electron.

A movable probe showed the existence of standing - wave patterns of the
oscillatory energy in the region of the plasma. Nodes of the patterns
coincided with the electrodes which limited the region of the plasma
traversed by the beam. Each of the observed frequency plateau corresponded
to the definite number of waves in a standing - wave pattern.

We argue that Looney and Brown have observed nonlinear plasma oscillations
stabilized by the plasma emission. It was shown recently ([5] and Refs
therein) that thermal emission has a stimulated character. According to this
conception thermal emission from non-uniform gas is produced by an ensemble
of individual emitters. Each of these emitters is an elementary resonator
the size of which has an order of magnitude of mean free path \textit{l} of
photons

\begin{equation}
\label{eq2}
l = \frac{{1}}{{n\sigma} }
\end{equation}

\noindent
where \textit{n} is the number density of particles and $\sigma $ is the
absorption cross-section.

\begin{center}
The emission of each elementary resonator is coherent, with the wavelength
\end{center}

\begin{equation}
\label{eq3}
\lambda = al,
\end{equation}

\noindent
where \textit{a} is a dimensionless constant, and thermal emission of
gaseous layer is incoherent sum of radiation produced by individual
emitters.

An elementary resonator emits in the direction opposite to the direction of
the density gradient. The wall of the resonator corresponding to the lower
density is half-transparent due to the decrease of absorption with the
decreasing gas density.

The relation (\ref{eq3}) implies that the radiation with the wavelength $\lambda $
is produced by the gaseous layer with the definite number density of
particles \textit{n} .

In the case of the experiment carried out by Looney and Brown, $n = n_{i} =
n_{e} $, due to the neutrality of the plasma, and the absorption
cross-section we assume to be equal to those for the fully ionized plasma,
$\sigma = \sigma _{0} = 10^{ - 9}cm^{2}$. The wavelength of nonlinear plasma
wave coincides with the wavelength of plasma emission as determined by the
relation (\ref{eq3}). The wavelength, $\lambda $, of standing wave can be calculated
as $\lambda = d/N$, where $d = 1.5cm$is the plasma length, the thickness of
the ion sheaths at each of two electrodes which limited the plasma region
traversed by the electron beam being neglected, and \textit{N} is the number
of waves in a standing - wave pattern. The electron density corresponding to
the strongest signal for each frequency plateau can be taken from Fig.2 in
[4].

From equations (\ref{eq2}) and (\ref{eq3}), we find the wavelength of standing wave in the
form

\begin{equation}
\label{eq4}
\lambda = a/\left( {n_{e} \sigma _{0}}  \right).
\end{equation}

Since $\lambda = d/N$, where N is a half-integer number,
oscillations would be observed only at definite values of the
electron density $n_{e} $. However, an elementary resonator has a
finite bandwidth [6]

\begin{equation}
\label{eq5}
\Delta \nu \approx v_{Te} /al,
\end{equation}

\noindent
where $v_{Te} $ is the thermal velocity of electrons. Due to the finite
bandwidth of an elementary resonator, oscillations can be observed also at
other values of the electron density, the intensity of the oscillation at
the ``wings'' of bandwidth much lower than the resonant intensity, as it was
indeed observed by Looney and Brown.

According to the relation (\ref{eq4}), the product of the wavelength, $\lambda $, by
the electron density, $n_{e} $, corresponding to the maximum intensity of
oscillations, should be a constant for all frequency plateaus observed by
Looney and Brown. In Table 1 we give the calculated values of $a = \lambda
n_{e} \sigma _{0} $ for each frequency plateau.

\begin{center}
\textbf{Table 1}
\end{center}

\begin{center}
An elementary resonator in the Looney - Brown experiment.
\end{center}

\newcommand{\PreserveBackslash}[1]{\let\temp=\\#1\let\\=\temp}
\let\PBS=\PreserveBackslash
\begin{longtable}
{|p{99pt}|p{85pt}|p{85pt}|p{85pt}|}
a & a & a & a  \kill
\hline
Number of waves \par \textit{N}&
Wavelength \par \[
\lambda
\]
&
Electron density \par \[
n_{e} \left( {10^{9}cm^{ - 3}} \right)
\]
&
Product \par \[
a = \lambda n_{e} \sigma _{0}
\]
 \\
\hline
\endhead
3/2&
1.0&
4.0&
4.0 \\
\hline
2&
0.75&
6.0&
4.5 \\
\hline
5/2&
0.6&
7.5&
4.5 \\
\hline
3&
0.5&
10.0&
5.0 \\
\hline
7/2&
0.4&
12.5&
5.0 \\
\hline
\end{longtable}

The wavelength, $\lambda $, of standing wave is calculated as $\lambda =
d/N$, where $d = 1.5cm$is the plasma length, and \textit{N} is the number of
waves in a standing - wave pattern. The electron density, $n_{e} $,
corresponding to the maximum intensity of oscillations, is taken from Fig.2
in [4]. The absorption cross-section is assumed to be equal $\sigma _{0} =
10^{ - 9}cm^{2}$.

The calculated value of \textit{a} is slightly varying with the electron
density $n_{e} $. A possible explanation is that the relation (\ref{eq4}) has a
statistical origin and suggests a sufficiently large number of waves,
\textit{N}. However, in the experiment by Looney and Brown, only the lowest
modes with small values of \textit{N} were observed. For small numbers of
waves in a standing - wave pattern, the edge effects, such as the ion
sheaths at each of two electrodes which limited the plasma region traversed
by the electron beam, are essential and can disturb the experimental
results. An accurate estimation of such edge effects with account for the
interaction of a plasma with radiation seems to be a difficult problem.

There seem to be different types of envelope solitary waves stabilized by
radiation. One of these types of radiation-induced solitary waves in a
plasma may be called a Looney-Brown soliton. It is an ordinary Langmuir
solitary wave [3,7] in which the radiation pressure is small compared to the
high-frequency pressure, but the wavelength of a carrying mode is determined
by the relation (\ref{eq4}), similar to the wavelength of the periodic wave in the
experiment by Looney and Brown.

The periodic wave is unstable with respect to the formation of local regions
of high-frequency electric field. Plasma is pushed out from these regions by
high-frequency pressure, so that the cavities of lower plasma density are
formed, an electric field being locked in these cavities. The Looney-Brown
envelope soliton represents such a localized region of lower plasma density
travelling at a group velocity of electron plasma waves

\begin{equation}
\label{eq6}
u = 3T_{e} k_{0} /\left( {m_{e} \omega _{p0}}  \right) \quad ,
\end{equation}

\noindent
where $T_{e} $ is the electron temperature, the electron plasma frequency
$\omega _{p0} $ corresponds to the undisturbed plasma density $n_{0} $, and
$k_{0} $ is the wave number of the carrying mode,

\begin{equation}
\label{eq7}
k_{0} = 2\pi /\lambda = 2\pi n_{0} \sigma _{0} /a.
\end{equation}

Here we make use of the equation (\ref{eq4}) for the wavelength $\lambda $.
Substituting relations (\ref{eq1}) and (\ref{eq7}) in equation (\ref{eq6}), we obtain

\begin{equation}
\label{eq8}
u = \left( {9\pi T_{e}^{2} n_{0} \sigma _{0}^{2} /am_{e} e^{2}}
\right)^{1/2}.
\end{equation}

For the values of the electron temperature $T_{e} \cong 1eV$ and the plasma
density $n_{0} \cong 10^{12}cm^{ - 3}$ which are characteristic for
low-temperature high-density plasma, the last equation gives $u \cong 3
\times 10^{8}cms^{ - 1}$.

Another type of radiation-induced solitary wave is the density waves
propagating in hot plasmas of accretion disks surrounding the compact
astrophysical objects such as active galactic nuclei and pulsars [6]. In
this case, the radiation pressure is not small and cannot be neglected.

\newpage

\begin{center}
----------------------------------------------------------------------------------------
\end{center}

[1] I.D.Kaganovich, Phys. Rev. Lett. \textbf{82}, 327 (1999).

[2] F.F.Chen, \textit{Introduction to Plasma Physics and Controlled Fusion,
Vol. 1: Plasma Physics} (Plenum Press, New York, 1984).

[3] V.E.Zakharov, in \textit{Handbook of Plasma Physics}, eds.
M.N.Rosenbluth and R.Z.Sagdeev, \textit{Vol. 2: Basic Plasma Physics II},
ed. by A.A.Galeev and R.N.Sudan (North- Holland Physics Publishers,
Amsterdam, 1984).

[4] D.H.Looney and S.C.Brown, Phys. Rev. \textbf{93}, 965 (1954).

[5] F.V.Prigara, Astron. Nachr\textit{.}, \textbf{324}, No. S1, 425 (2003).

[6] F.V.Prigara, Phys. Plasmas (submitted), E-print archives,
physics/0404087.

[7] B.B.Kadomtsev, \textit{Collective Phenomena in Plasmas} (Nauka, Moscow,
1988).

\end{document}